\documentclass[twocolumn,showpacs,superscriptaddress,preprintnumbers,prl,amsmath,amssymb,floatfix,showkeys]{revtex4}

\usepackage{graphicx}
\usepackage{bm}
\begin{document}

\title{Muon spin relaxation and isotropic pairing in superconducting $\mathbf{PrOs_4Sb_{12}}$}

\author{D. E. MacLaughlin}
\affiliation{Department of Physics, University of California, Riverside, California 92521-0413}

\author{J. E. Sonier}
\affiliation{Department of Physics, Simon Fraser University, Burnaby, B.C., Canada \mbox{V5A 1S6}}
\affiliation{Canadian Institute for Advanced Research, Toronto, Ontario, Canada \mbox{M5G 1Z8}}

\author{R. H. Heffner}
\affiliation{MS K764, Los Alamos National Laboratory, Los Alamos, New Mexico 87545}

\author{O. O. Bernal}
\affiliation{Department of Physics and Astronomy, California State University, \\Los Angeles, California 90032}

\author{Ben-Li Young}
\author{\\M. S. Rose}
\affiliation{Department of Physics, University of California, Riverside, California 92521-0413}

\author{G. D. Morris}
\affiliation{MS K764, Los Alamos National Laboratory, Los Alamos, New Mexico 87545}

\author{E. D. Bauer}
\author{T. D. Do}
\author{M. B. Maple}
\affiliation{Department of Physics, University of California, San Diego,\\ La Jolla, California 92093-0319}

\date{\today}

\begin{abstract}
Transverse-field muon spin rotation measurements in the vortex lattice of the heavy-fermion (HF) superconductor $\mathrm{PrOs_4Sb_{12}}$ yield a temperature dependence of the magnetic penetration depth $\lambda$ indicative of an isotropic or nearly isotropic energy gap. This is not seen to date in any other HF superconductor, and is a signature of isotropic pairing symmetry, possibly related to a novel nonmagnetic ``quadrupolar Kondo'' HF mechanism in $\mathrm{PrOs_4Sb_{12}}$. The $T = 0$ relaxation rate~$\sigma_s(0) = 0.91(1)~\mathrm{\mu s^{-1}}$ yields an estimated magnetic penetration depth~$\lambda(0) = 3440(20)$~\AA, which is considerably shorter than in other HF superconductors.
\end{abstract}

\maketitle

In most heavy-fermion (HF) metals and superconductors the $f$ ion $\mathrm{(Ce, Yb, U)}$ has a magnetic ground state. HF behavior has, however, been reported in a small number of praseodymium-based alloys and compounds~\cite{LeHa75,YBMC96,SAOM00}, in which the crystalline electric field (CEF) ground state of the non-Kramers $\mathrm{Pr^{3+}}$ ion could be nonmagnetic and degenerate. In this case a charge-scattering analog of Kondo spin scattering can give rise to the so-called ``quadrupolar Kondo effect''~\cite{Cox87}, an example of the two-channel Kondo effect that has been invoked to explain non-Fermi-liquid behavior in HF systems~\cite{CoJa96}.

Superconductivity has recently been discovered in the cubic HF compound~$\mathrm{PrOs_4Sb_{12}}$~\cite{BFHZ02,MHZF02}. From thermodynamic measurements it is found that a large carrier effective mass~$m^{\ast} \approx 50\,m_e$ characterizes both the normal and superconducting states, and the transition temperature~$T_c = 1.85$~K is relatively high for a HF superconductor. Although a conventional spin-based HF mechanism has not been completely ruled out, thermodynamic properties of $\mathrm{PrOs_4Sb_{12}}$ suggest a $\mathrm{Pr^{3+}}$ nonmagnetic doublet $\Gamma_3$ CEF ground state, so that the quadrupolar Kondo effect is a candidate mechanism for the HF behavior. The symmetry of the superconducting pairing in such a system is a fundamental question.

Transverse-field muon spin rotation (TF-$\mu$SR) has proved an effective probe of the internal magnetic field distribution~$n(B)$ in the vortex state of conventional and unconventional type-II superconductors~\cite{HeNo96,Amat97,SBK00}. In the $\mu$SR technique~\cite{Sche85,Brew94} spin-polarized positive muons ($\mu^+$) are implanted into the sample, where the muon spins precess in their local fields. In general the muon spin relaxation rate is related to the rms width~$[\overline{(\Delta B)^2}]^{1/2}$ of $n(B)$. In turn $[\overline{(\Delta B)^2}]^{1/2}$ is inversely proportional to the square of the magnetic penetration depth~$\lambda$, which is related to the density~$n_s$ of superconducting carriers and $m^{\ast}$ by the London equation
\begin{equation}
1/\lambda^2 = 4\pi n_s e^2/m^{\ast}c^2 \,.
\label{eq:London}
\end{equation}
The temperature dependence of $\lambda$ at low temperatures is therefore sensitive to the lowest-lying superconducting excitations, the thermal population of which reduces $n_s$ with increasing temperature. In the presence of an isotropic or nearly isotropic energy gap $\lambda(T) - \lambda(0)$ varies exponentially with temperature, whereas nodes in the gap function characteristic of non-$s$-wave pairing lead to power-law dependences~\cite{HeNo96,Amat97,SBK00}. In HF superconductors $\lambda$ is typically very long ($\gtrsim 10^4$ \AA~\cite{HeNo96}), in which case the muon relaxation rate can be dominated by other sources of magnetism in the sample. It is then difficult to extract $\lambda(T)$ from TF-$\mu$SR experiments, but the technique has been successfully applied in a number of systems~\cite{Amat97}.

This Letter reports $\mu$SR experiments in the superconducting state of $\mathrm{PrOs_4Sb_{12}}$. The penetration depth derived from the vortex-lattice field distribution width exhibits the temperature dependence characteristic of isotropic $s$-wave pairing. Isotropic $p$-wave pairing~\cite{BaWe63}, which is indistinguishable from $s$-wave pairing by thermodynamic or electrodynamic measurements, is also possible.  To our knowledge this is the only example to date of an isotropic gap in a HF superconductor. It suggests the possibility of (a)~marked differences in superconducting properties between HF materials with magnetic and nonmagnetic $f$-ion ground states, and (b)~a relation between the pairing symmetry and the mechanism (spin or quadrupole Kondo effect) for the HF normal state from which the superconductivity evolves.

$\mu$SR measurements were carried out in the dilution refrigerator at the M15 beam line, \mbox{TRIUMF}, Vancouver, Canada, on an unaligned powder sample of $\mathrm{PrOs_4Sb_{12}}$ prepared by growth from an antimony flux as described previously~\cite{BFHZ02,MHZF02}. Gallium arsenide, which rapidly depolarizes muons in transverse field, was placed around the sample to minimize any background signal due to muons that stop in the sample holder. TF-$\mu$SR and zero-field $\mu$SR (ZF-$\mu$SR) data were obtained for temperatures between 0.05 and 2.1~K\@.

In a number of HF superconductors local fields due to spontaneous static magnetism below $T_c$ are inferred from ZF-$\mu$SR data~\cite{HeNo96,Amat97}. Such magnetism can originate from a spin-wave instability or local-moment ordering, as observed in CeCu$_{2.2}$Si$_2$~\cite{FAGG97} where magnetic ordering and superconductivity compete for the ground state. Alternatively, static magnetism below $T_c$ may be due to a superconducting order parameter that breaks time-reversal symmetry, as observed in, e.g., Th-doped $\mathrm{UBe_{13}}$~\cite{HeNo96,HSWB90}. Muon local fields due to static magnetism could masquerade as vortex lattice inhomogeneity, and falsify the interpretation of the TF-$\mu$SR results given above.

\begin{figure}[ht]
\includegraphics[clip=,width=0.86\columnwidth]{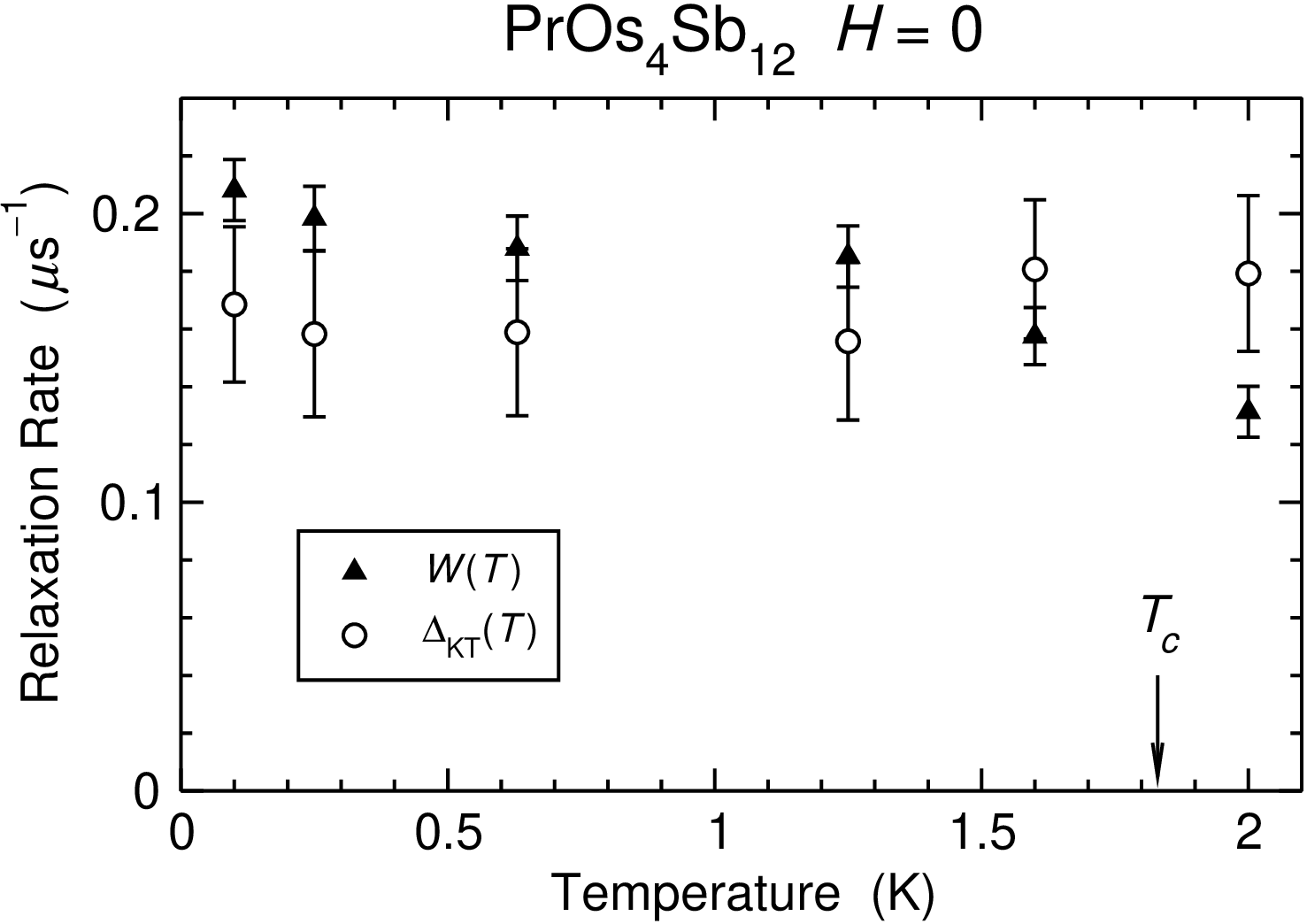}
\caption{Damped Kubo-Toyabe ZF-$\mu$SR relaxation in $\mathrm{PrOs_4Sb_{12}}$. Circles: Kubo-Toyabe relaxation rate~$\Delta_{\mathrm{KT}}(T)$. Triangles: exponential damping rate~$W(T)$.}
\label{fig:zf}
\end{figure}

We have therefore carried out ZF-$\mu$SR experiments to determine whether such static magnetism exists in $\mathrm{PrOs_4Sb_{12}}$. The zero-field data are well fit by a ``damped Kubo-Toyabe'' function, i.e., the product of a damping exponential and the Kubo-Toyabe function~\cite{HUIN79} expected from nuclear dipolar fields. Figure~\ref{fig:zf} gives the temperature dependence of the corresponding relaxation rates $W$ and $\Delta_{\mathrm{KT}}$.
The exponential damping rate~$W(T)$ is already appreciable in the normal state, increases only slightly more than experimental uncertainty below $T_c$, and is (negatively) correlated with $\Delta_{\mathrm{KT}}(T)$ in the fitting process. Thus there is no statistically significant evidence for static magnetism below $T_c$; the present data place an upper limit of $\sim 50~\mu$T on any static field below $T_c$. We argue below that the relaxation rate changes of Fig.~\ref{fig:zf} are too small to affect the considerably increased TF-$\mu$SR rates in the superconducting state.

\begin{figure}[ht]
\includegraphics[clip=,width=0.86\columnwidth]{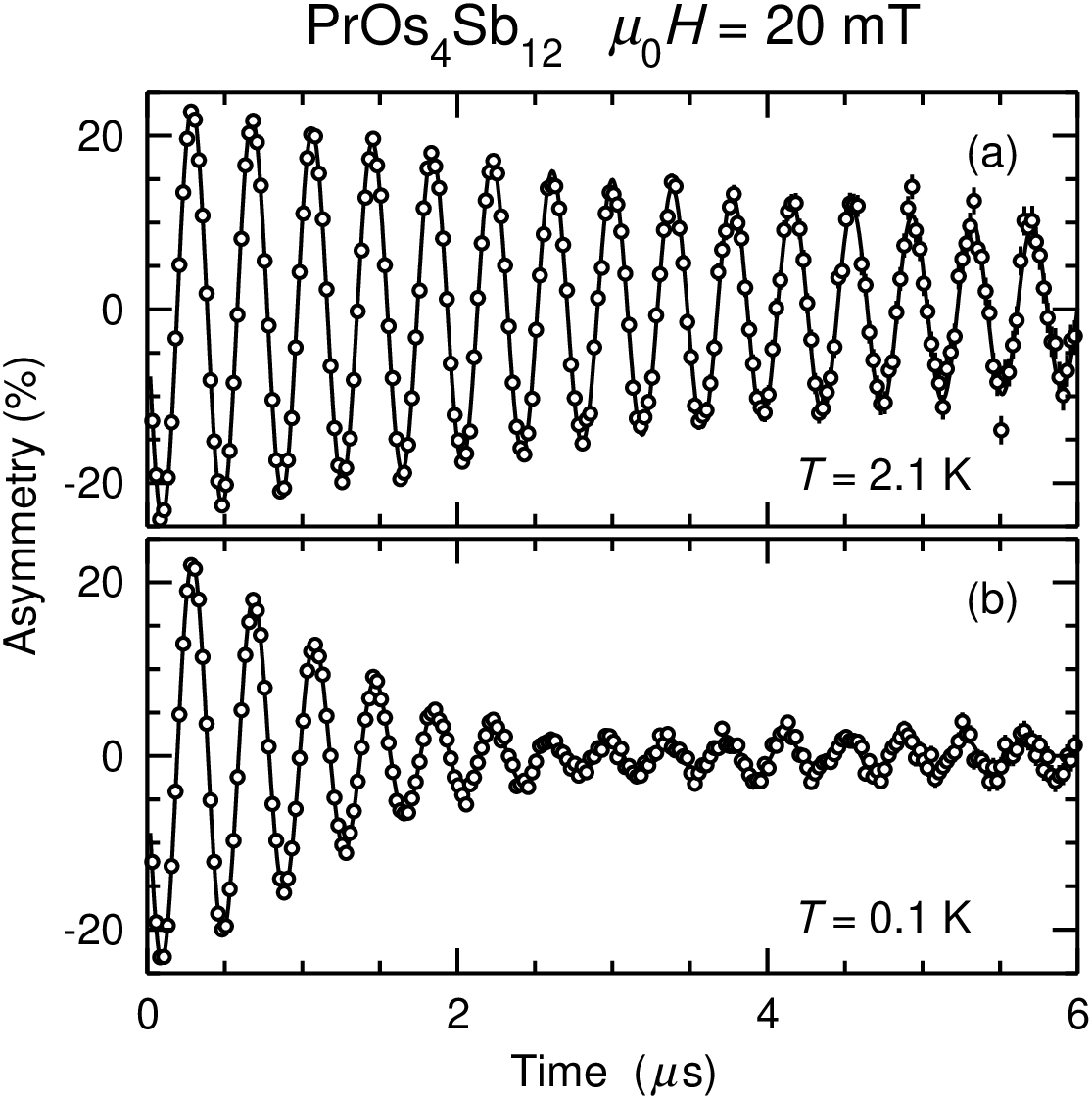}
\caption{TF-$\mu$SR spin precession signals in $\mathrm{PrOs_4Sb_{12}}$ ($T_c = 1.82$~K), applied field 20~mT. (a)~Normal state ($T = 2.1$~K)\@. (b)~Superconducting state ($T = 0.1$~K)\@. The weak nonrelaxing signal in (b) is due to muons that do not stop in the sample.}
\label{fig:asymm}
\end{figure}

Representative TF-$\mu$SR muon-spin precession signals at an applied field of 20~mT are shown in Fig.~\ref{fig:asymm} in the normal and superconducting states.
A small ($\sim$10\%) nonrelaxing background signal is visible at long times in Fig.~\ref{fig:asymm}(b). Normal-state data for temperatures just above $T_c$ are best fit by an exponential decay function, with a rate constant~$W_n = 0.17(1)~\mu\mathrm{s^{-1}}$ at 20~mT, rather than the Gaussian commonly found when the broadening is due to nuclear dipolar fields~\cite{SBK00}. This suggests that the broadening is due to a different mechanism, most likely inhomogeneity in the Pr-moment susceptibility. Consistent with this view, in the normal state $W_n$ grows linearly with applied field (data not shown).

The internal field distribution in the vortex state is the convolution of the field distributions due to the vortex lattice and to the electronic and nuclear moments of the host material; by the convolution theorem the muon-spin precession signal is the product of the Fourier transforms of these distributions. The superconducting-state data have been fit to a product of exponential and Gaussian functions, where the latter is taken as an approximation to the relaxation function due to the vortex-lattice field distribution. The exponential rate~$W_n$ was held fixed at the normal-state value, although at low temperatures the values of the superconducting-state Gaussian rate~$\sigma_s$ are not greatly changed if $W_n$ is allowed to vary for best fit.

The fit is statistically satisfactory, as can be seen qualitatively in Fig.~\ref{fig:asymm}(b). This indicates that $n(B)$ is essentially Gaussian, due perhaps to disorder in the vortex lattice; the non-Gaussian relaxation expected from the field distribution in an ideal vortex lattice~\cite{SBK00,Bran88} was not observed. The Gaussian relaxation rate~$\sigma_s$ is then given by $\gamma_\mu [\overline{(\Delta B)^2}]^{1/2}$, where $\gamma_\mu = 8.516 \times 10^8~\mathrm{s^{-1}~T^{-1}}$ is the $\mu^+$ gyromagnetic ratio.

\begin{figure}[ht]
\includegraphics[clip=,width=0.86\columnwidth]{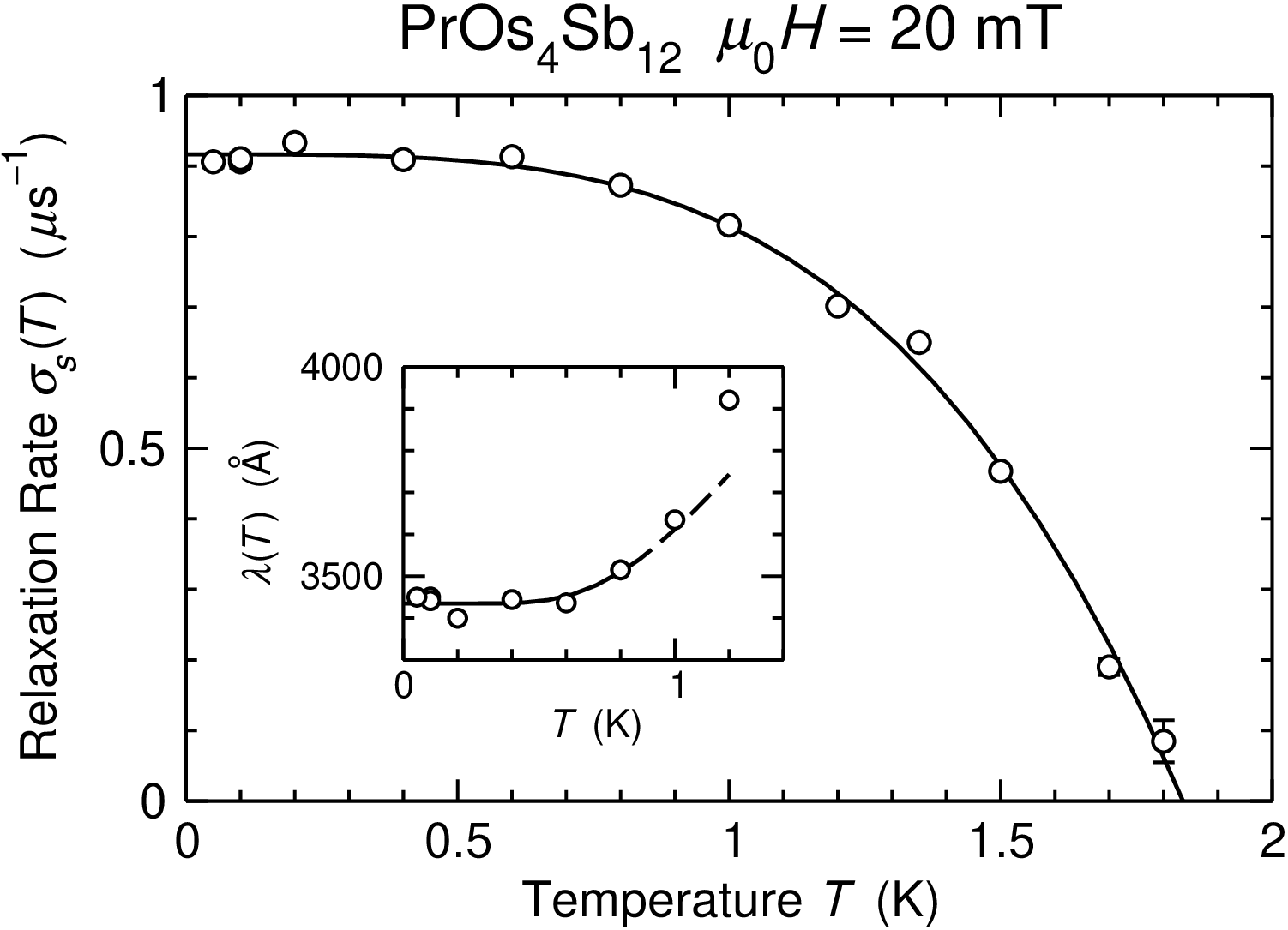}
\caption{Temperature dependence of vortex-state $\mu^+$ relaxation rate~$\sigma_s(T)$ in the superconducting state of $\mathrm{PrOs_4Sb_{12}}$. Curve: $\sigma_s(T) = \sigma_s(0)[1-(T/T_c)^y]$, $\sigma_s(0) = 0.91(1)~\mu{\mathrm s}^{-1}$, $T_c = 1.83(2)$~K, $y = 3.6(2)$. Inset: low-temperature penetration depth $\lambda(T)$ derived from $\sigma_s(T)$. Curve: $\lambda(T) = \lambda(0)[1 + (\pi\Delta/2T)^{1/2}\exp(-\Delta/T)]$, $\lambda(0) = 3440(20)$~\AA, $\Delta/T_c = 2.1(2)$, from fit to data for $T \le 0.8$~K.}
\label{fig:s1}
\end{figure}

The temperature dependence of $\sigma_s$ is shown in Fig.~\ref{fig:s1} for an applied field of 20~mT.
Salient features of these data are that $\sigma_s(T)$ varies only slowly at low temperatures, and that the $T = 0$ value $\sigma_s(0) = 0.91(1)~\mu{\mathrm s}^{-1}$. This is large for a HF superconductor~\cite{HeNo96,Amat97} and is much larger than changes in the ZF-$\mu$SR rates (Fig.~\ref{fig:zf}), so that $\sigma_s$ is dominated by the vortex-state field distribution.

We obtain $\lambda$ from $\sigma_s$ and the expression
\begin{equation}
\overline{(\Delta B)^2} = 0.00371\, \Phi_0^2\lambda^{-4}
\label{eq:Brandt}
\end{equation}
appropriate to an isotropic extreme type-II superconductor~\cite{Bran88}, where $\Phi_0$ is the flux quantum. The data are shown in the inset to Fig.~\ref{fig:s1}. The BCS low-temperature expression~$\lambda(T) = \lambda(0)[1 + (\pi\Delta/2T)^{1/2}\exp(-\Delta/T)]$~\cite{Mueh59} was fit to data for $T \le 0.8~\mathrm{K} \lesssim 0.4\,T_c$; the curve in the inset to Fig.~\ref{fig:s1} shows this fit and its extension (dashed) up to 1.2 K\@. The fit value~$\lambda(0) = 3440(20)$~\AA\ is short for a HF superconductor. The fit value of the ratio~$\Delta/T_c = 2.1(2)$ is somewhat larger than the BCS value~1.76, suggesting strong coupling. Over the entire temperature range $\sigma_s(T)$ is consistent with the phenomenological ``two-fluid'' temperature dependence~$1/\lambda^2(T) \propto 1-(T/T_c)^4$, although the data are slightly better fit with an exponent of $3.6(2)$ (curve). All these properties indicate that the gap is isotropic or nearly so~\cite{HeNo96,Amat97,SBK00}. Proportionality between $\overline{(\Delta B)^2}$ and $\lambda^{-4}$  should survive vortex lattice disorder as long as the distance between vortices is much smaller than $\lambda$ ($H \gg H_{c1}$); disorder increases the numerical coefficient in Eq.~(\ref{eq:Brandt})~\cite{Bran88} but should not affect the temperature dependence of $\overline{(\Delta B)^2}$.

Early studies of the penetration depth in cuprate superconductors using unaligned powders also observed little temperature dependence of $\lambda$ at low temperatures, and concluded that the pairing was $s$-wave~\cite{SBK00}. Later measurements in high-quality aligned crystals~\cite{SKBB74} revealed the linear temperature dependence characteristic of $d$-wave pairing. The early results appear to have been due to a combination of circumstances: strong anisotropy in $\lambda$, oxygen inhomogeneity, and sensitivity to hole doping due to the low density~$n_s$ of superconducting carriers. None of these factors would seem to affect the current measurements. The crystal structure of $\mathrm{PrOs_4Sb_{12}}$ is cubic, so that the penetration depth is isotropic. With $m^{\ast} \approx 50\,m_e$ we find $n_s \approx 10^{22}~\mathrm{carriers/cm^3}$ from Eq.~(\ref{eq:London}), so that $\mathrm{PrOs_4Sb_{12}}$ is a good metal and the carrier concentration is insensitive to chemical inhomogeneity. Thus it seems unlikely that the observed temperature dependence of $\lambda(T)$ is due to extrinsic effects.

The resistive mean free path~$\ell$ can be calculated from the residual resistivity~$\rho(0) \approx 5~\mu\Omega$-cm~\cite{BFHZ02,MHZF02} using $n_s$ as an estimate of the normal-state carrier concentration, and can be compared with the superconducting coherence length~$\xi_0$~\cite{BFHZ02}. We find $\ell/\xi_0 \approx 3$, so that $\mathrm{PrOs_4Sb_{12}}$ is a rather clean superconductor [and Eq.~(\ref{eq:London}), derived in the clean limit, is valid]; this is of importance in the analysis of thermodynamic properties.

\begin{figure}[ht]
\includegraphics[clip=,width=0.86\columnwidth]{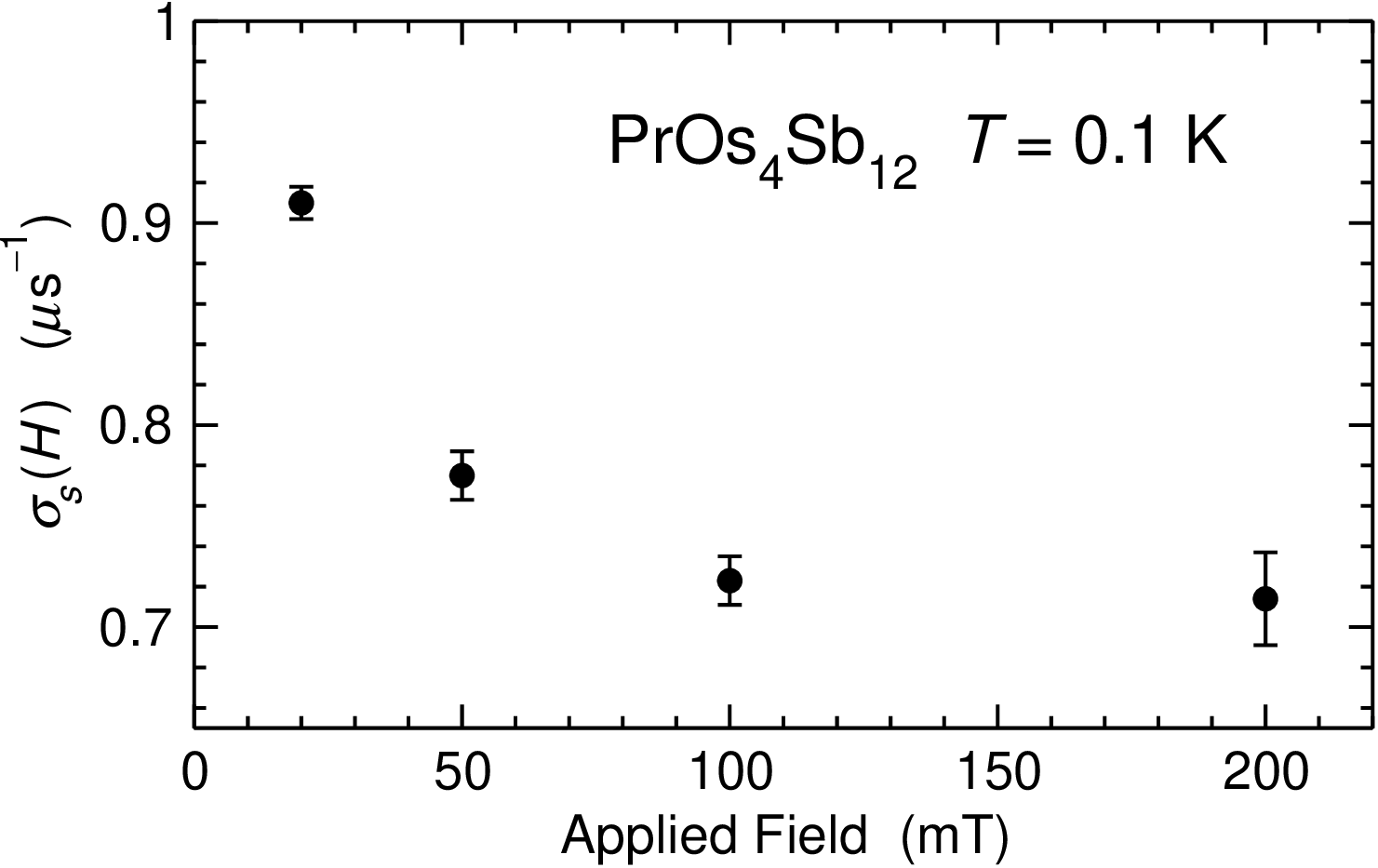}
\caption{Dependence of TF-$\mu$SR relaxation rate $\sigma_s(H)$ on  applied field in $\mathrm{PrOs_4Sb_{12}}$, $T = 0.1$~K.}
\label{fig:svsH}
\end{figure}

The field dependence of $\sigma_s$ at $T = 0.1$~K in $\mathrm{PrOs_4Sb_{12}}$ is shown in Fig.~\ref{fig:svsH}.
In these experiments the field was set with the temperature above $T_c$, and the sample was cooled into the superconducting state at constant field; otherwise trapped flux broadens the field distribution and causes a spurious increase of the $\mu^+$ relaxation rate. The field dependence of $\sigma_s$ is nonlinear, decreasing at low fields and then saturating above $\sim 100$~mT\@. The origin of this nonlinearity is not understood, but might be due to disorder in the vortex lattice~\cite{SBK00,Bran88} or to a change of vortex-lattice symmetry, e.g., from a triangular to a square lattice. Measurements at fields higher than $\sim 200$~mT will be difficult because of the growing importance of the normal-state paramagnetic relaxation rate.

\begin{figure}[ht]
\includegraphics[clip=,width=0.86\columnwidth]{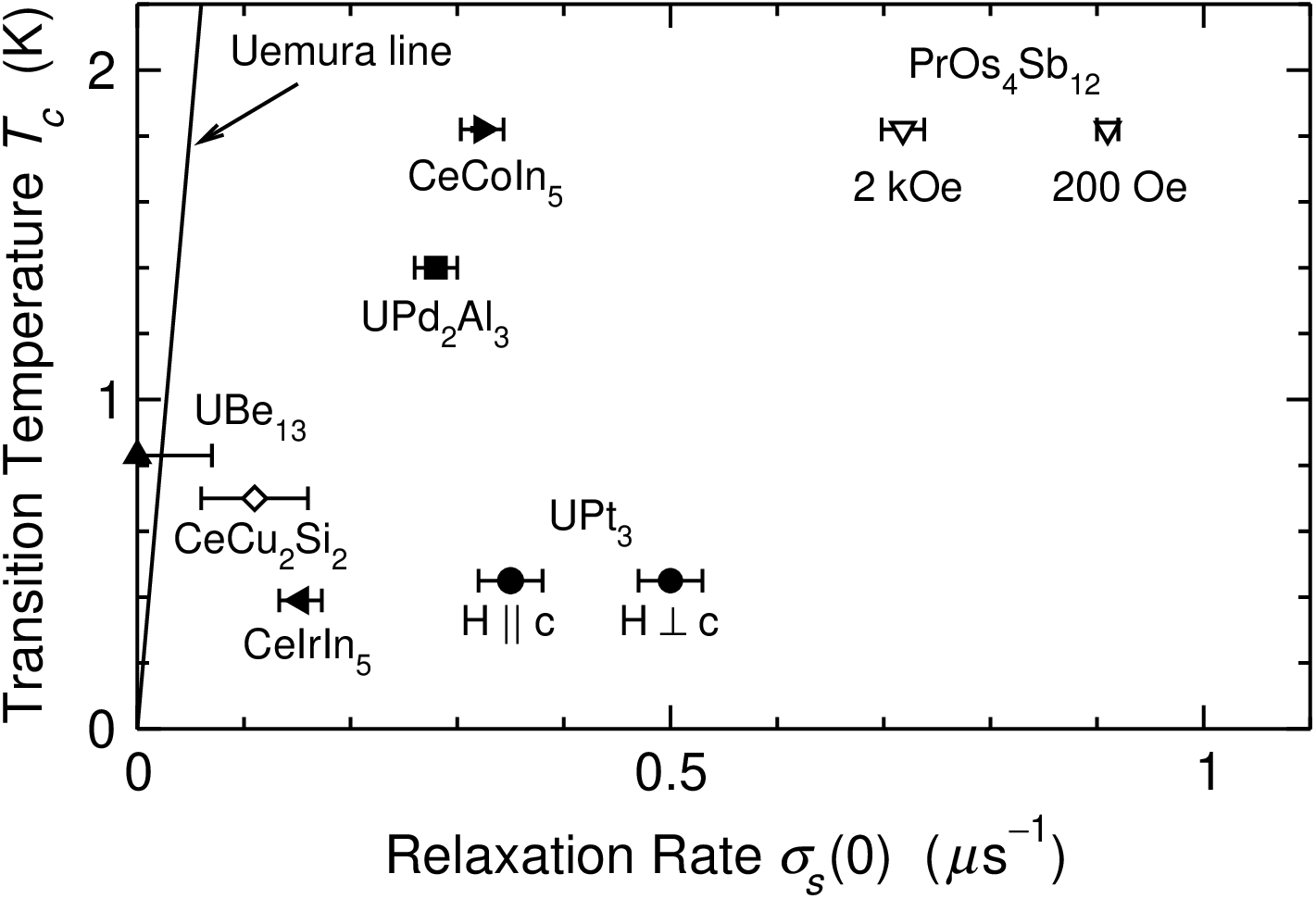}
\caption{Uemura plot of superconducting transition temperature~$T_c$ versus muon relaxation rate~$\sigma_s(0)$ for $\mathrm{UBe_{13}}$ \protect\cite{DYHS00}, $\mathrm{CeCu_2Si_2}$ \protect\cite{Amat97}, $\mathrm{UPd_2Al_3}$ \protect\cite{FAGS94b}, $\mathrm{Ce}T\mathrm{In_5}$, $T = \mathrm{Co}$ and Ir \protect\cite{HKKK02}, $\mathrm{UPt_3}$ \protect\cite{YDHF98}, and $\mathrm{PrOs_4Sb_{12}}$ (this Letter). The ``Uemura line'' is followed by a number of unconventional superconductors.}
\label{fig:Uemura}
\end{figure}

Figure~\ref{fig:Uemura} is an Uemura plot~\cite{SBK00,ULLS91} of superconducting transition temperature~$T_c$ vs zero-temperature muon relaxation rate~$\sigma_s(0)$ in a number of HF superconductors: $\mathrm{UBe_{13}}$~\cite{DYHS00}, $\mathrm{CeCu_2Si_2}$ (derived from the penetration depth estimate of Ref.~\cite{Amat97}), $\mathrm{UPd_2Al_3}$~\cite{FAGS94b}, $\mathrm{CeCoIn_5}$ and $\mathrm{CeIrIn_5}$,~\cite{HKKK02}, $\mathrm{UPt_3}$~\cite{YDHF98}, and $\mathrm{PrOs_4Sb_{12}}$ (this Letter).
It can be seen that $\sigma_s(0)$ is larger in $\mathrm{PrOs_4Sb_{12}}$ than in other HF superconductors. The values of $\sigma_s(0)$ shown in Fig.~\ref{fig:Uemura} are in rough inverse proportion to the HF effective masses of these systems, as expected from Eq.~(\ref{eq:London}) and $\sigma_s(0) \propto 1/\lambda^{2}(0)$, so that the position of $\mathrm{PrOs_4Sb_{12}}$ in this plot is not surprising.

The isotropy of the superconducting energy gap indicated by the temperature dependence of $\sigma_s$ (Fig.~\ref{fig:s1}) strongly suggests that superconductivity in $\mathrm{PrOs_4Sb_{12}}$ is in some sense ``conventional.'' In this regard $\mathrm{PrOs_4Sb_{12}}$ differs markedly from other HF superconductors studied to date, a difference that may possibly be related to a nonmagnetic or quadrupolar Kondo state in this compound. Better understanding of the quadrupolar Kondo lattice will be needed to elucidate the questions of whether $\mathrm{PrOs_4Sb_{12}}$ is indeed such a system and, if so, whether this property is related to the conventional superconducting behavior.

\begin{acknowledgments}
We are grateful to D. Arsenau, B. Hitti, and S. Kreitz\-man for help with these experiments, and to A. Amato and D. Cox for useful discussions. This work was supported in part by the U.S. NSF, Grant nos.~DMR-0102293 (UC Riverside), DMR-9820631 (CSU Los Angeles) and DMR-0072125 (UC San Diego), the Canadian NSERC (Simon Fraser), the U.S. DOE, Grant no.~DE-FG03-86ER-45230 (UC San Diego), and the NEDO International Joint Research Program (UC San Diego), and was performed in part under the auspices of the U.S. DOE (Los Alamos).
\end{acknowledgments}

\end{document}